\documentstyle[11pt,newpasp,twoside,epsf]{article}
\def\lsim {\lower .1ex\hbox{\rlap{\raise .6ex\hbox{\hskip .3ex
        {\ifmmode{\scriptscriptstyle <}\else
                {$\scriptscriptstyle <$}\fi}}}
        \kern -.4ex{\ifmmode{\scriptscriptstyle \sim}\else
                {$\scriptscriptstyle\sim$}\fi}}}
\def\eps@scaling{.95}
\def\epsscale#1{\gdef\eps@scaling{#1}}
 \def\plotone#1{\centering \leavevmode
\epsfxsize=\eps@scaling\textwidth \epsfbox{#1}}
\def\edcomment#1{\iffalse\marginpar{\raggedright\sl#1\/}\else\relax\fi}
\newcommand{\etal}{{et al.~}}
\newcommand{\Mo}{{\rm M_\odot}}
\reversemarginpar

\begin{document}
\title{Galaxies and Overmerging: What Does it Take to Destroy a 
Satellite Galaxy?}
\author{Stelios Kazantzidis, Ben Moore \& Lucio Mayer}
\affil{Institute for Theoretical Physics, University of Z\"urich, Switzerland}

\begin{abstract}

The Ultra Compact Dwarf (UCD) galaxies recently discovered in the
Fornax and Virgo clusters exhibit structural similarity to the dense nuclei 
of nucleated dEs indicating that the progenitor 
galaxy and its halo have been entirely tidally disrupted.
Using high resolution $N$-body simulations
with up to ten million particles we investigate the evolution and tidal 
stripping of substructure halos orbiting within a host potential.
We find that complete disruption of
satellite halos modeled following the NFW density profile occurs only 
for very low values of concentration in disagreement with the 
theoretical predictions of CDM models. This discrepancy is further
exacerbated when we include the effect of baryons since disk
formation increases the central density.
\end{abstract}

\section{Introduction}

High resolution numerical simulations and 
sophisticated semi-analytic modeling have
significantly improved our understanding of the properties and the 
evolution of cold dark matter (CDM) substructure. 
It has been demonstrated (Moore \etal 1998, Colpi, Mayer, \& Governato 1999; 
Moore \etal 1999; Klypin \etal 1999; Taffoni \etal 2003) that tidal distruption is very 
inefficient for these low mass subhalos. Satellites represent
earlier generations of the merging hierarchy and are typically denser
and more concentrated than their more massive hosts. 
While gravitational tides serve to unbind 
mass associated with these subhalos, it is still not clear whether or not 
complete disruption of substructure halos can take place 
in a CDM potential.

Direct evidence of tidal disruption processes operating
effectively within galaxy clusters has recently been provided 
by the discovery of a new population of subluminous and extremely 
compact objects in the Fornax (Drinkwater \etal 2000; Phillipps \etal 2001)
and Virgo (Drinkwater, private communication) clusters.
These Ultra Compact Dwarf (UCD) galaxies are 
dynamically distinct systems with intrinsic sizes $\lsim 100$\ pc,
and properties, including velocity dispersions, absolute magnitudes and
mass-to-light (M/L) ratios, considerably higher than any normal globular 
cluster (Drinkwater \etal 2003). There is accumulating evidence that the UCDs
constitute the remnant nuclei of dwarf galaxies whose extended stellar 
component and DM halo have been both entirely disrupted by gravitational 
interactions within their host cluster.
Their derived M/L ratios range from 2 to 4 and are consistent
with those of stellar populations suggesting that these systems contain
no dark matter. 
Other factors consistent with the interpretation that the UCDs are 
the products of tidal disruption processes include their 
strong structural similarity to the dense nuclei of nucleated dwarf ellipticals (dEs) 
and the lack of extended stellar envelopes around them in photographic images 
(Phillipps \etal 2001). 

Our goal is not only to investigate how probable
complete disruption of substructure halos at the typical distances of known 
UCDs (within about 30\% of the cluster virial radius) is, but also to place constraints 
on the structure of DM halos and examine whether the existence of the UCDs  
is consistent with the theoretical predictions of CDM models.

\section{Numerical Simulations}

We study the evolution of dwarf satellites comparable in mass to typical
cluster nucleated dEs in the external potential of the Fornax cluster.
The NFW density profile (Navarro, Frenk, \& White 1996)
is used for both the live satellites and the spherically symmetric static 
host potential. The latter represents a cluster halo 
with virial mass $M_{\rm prim}=0.5 \times10^{14} \,h^{-1} \Mo$ 
and $c_{\rm prim}=8.5$ (hereafter, $h=0.5$).
The satellite's virial mass is $M_{\rm sat}=2 \times10^{10} \,h^{-1} \Mo$.
Our simulations neglect the effects of dynamical friction 
and the response of the primary to 
the presence of the satellite. Considering the vast difference 
in the mass and size of the two main systems we do not expect our 
results to be affected by this choice.
In this investigation all satellite models are
Monte Carlo realizations of the exact phase-space distribution function 
under the assumptions of 
spherical symmetry and isotropic velocity dispersion tensors, $f=f(E)$.
Kazantzidis, Magorrian, \& Moore (2003) have explicitly demonstrated that 
the choice of initial conditions is vital for studies like the present.
Out of equilibrium initial conditions may artificially accelerate the mass loss
of the model satellites by decreasing their central density and changing 
the character of their orbital anisotropy.

The current positions of the UCDs which give an indication of the 
apocenter of their orbits coupled with 
theoretical studies of halo orbital properties, 
will be used to constrain the orbital parameters of the satellites.
In particular, we shall adopt an apocenter radius equal to 
$r_{\rm apo}=1.77\ R_{\rm s}$, 
where $R_{\rm s}$ is the scale radius of the host halo and 
$(r_{\rm apo}/r_{\rm peri})=\,$(5:1) close to the median ratio
of apocentric to pericentric radii found in cosmological simulations
(Ghigna \etal 1998).
The pericenter of the orbit is $50$ kpc in all the simulations presented here.
We evolve our models using PKDGRAV (Stadel 2001) for 10 Gyr.
In all our runs, the total energy was conserved to better than 0.1\%.
In Figure 1 (left panel) we perform a quantitative comparison 
of the evolution of the bound satellite mass for three 
different mass resolutions. We use the group finder SKID 
(Stadel 2001) to identify the remaining bound mass.
We define complete disruption of a satellite system when we do not
find any gravitationally bound structure at a scale of $2 \varepsilon_s$ or larger,
where $\varepsilon_s$ is the gravitational softening for our runs.
The satellite halo has a $c_{\rm sat}=5$ and is simulated 
with $N=10^5$ (filled squares), $N=10^6$ (open circles), 
and $N=10^7$ (filled circles) particles.
The evolution of the bound mass is plotted up to
the point where complete disruption of the satellite system occurs.
The halo resolved with just $10^5$ particles fully disrupts after just two orbits, 
but this is clearly a resolution effect since the same halos resolved 
with more particles survive significantly longer. 
Convergence is achieved when we adopt a mass resolution of 
more than $10^6$ particles per halo.

\begin{figure}
\plottwo{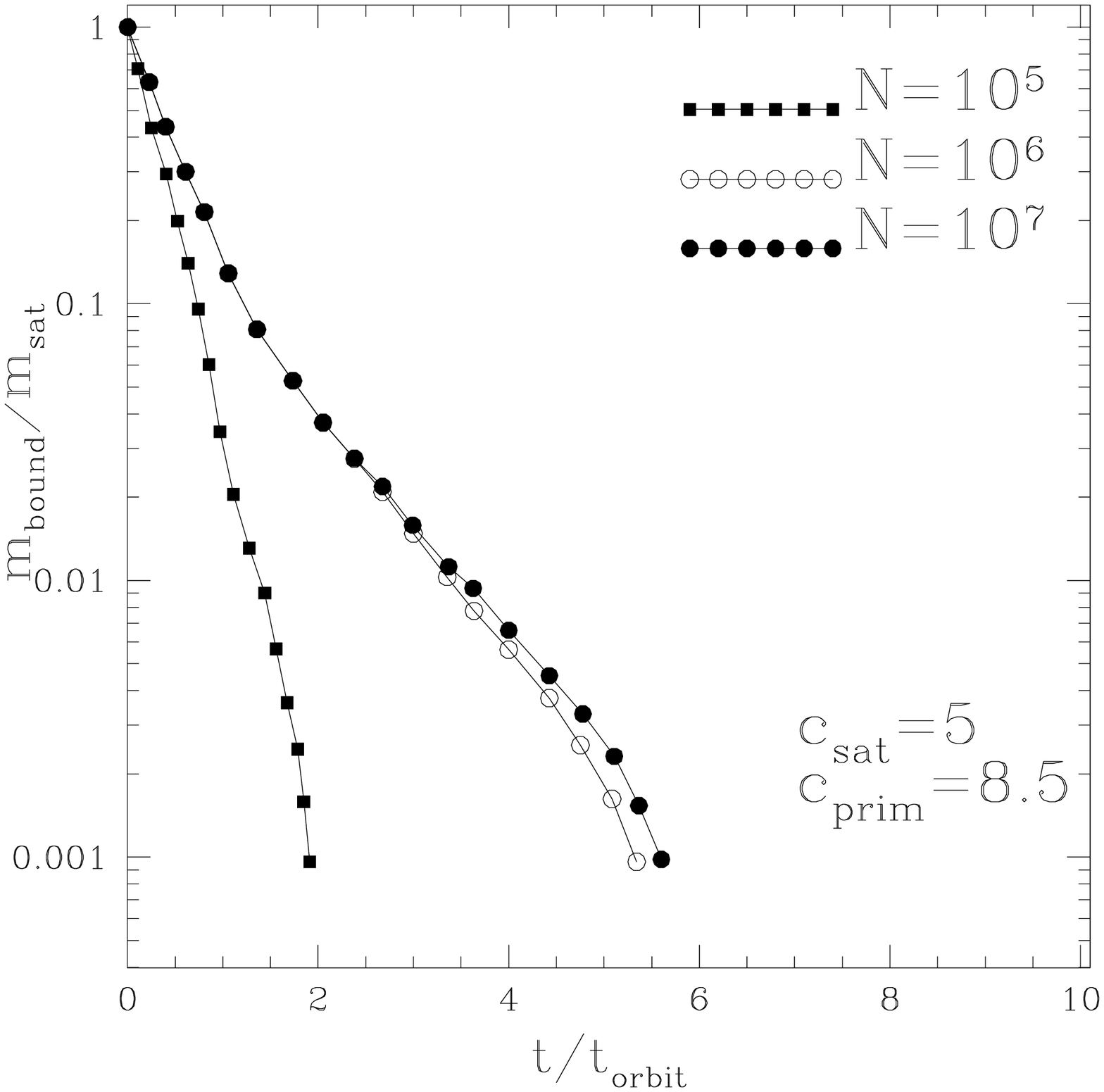}{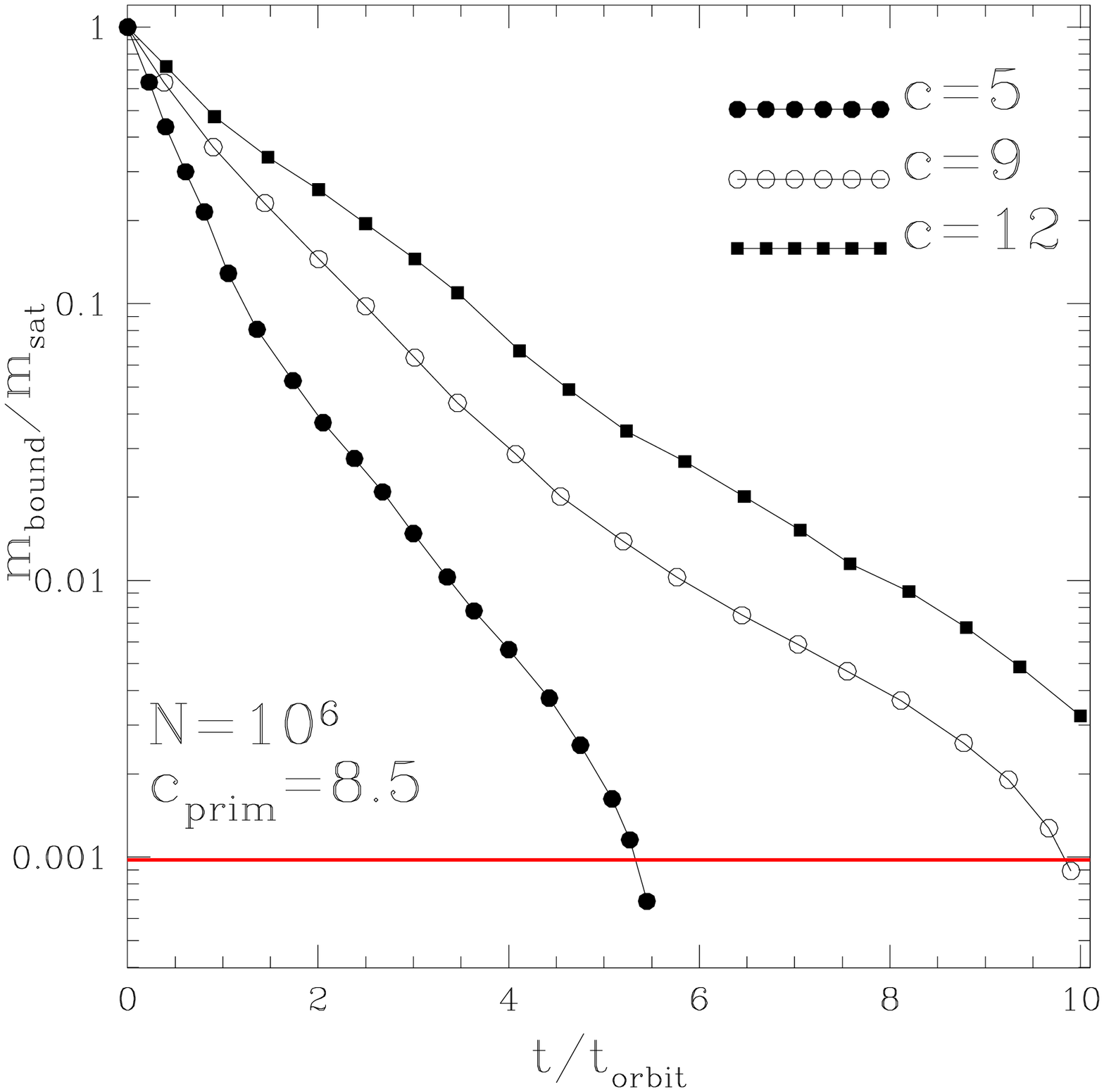}
\caption{Left: Bound satellite mass as a function of the orbital period
for three different mass resolutions. Results approach convergence when 
adopting a mass resolution of the order of a few million particles. 
Right: Evolution of the bound NFW satellite mass for three different 
concentrations. The value $c_{\rm sat}=9$ gives the upper limit 
for disruption at the timescales of interest.}
\label{fig1}
\end{figure}

In order to avoid an unecessary computational cost, 
we adopt a mass resolution of $N=10^6$ in the following runs.
We set the gravitational softening to $\epsilon=50$\ pc, 
hence, our force resolution being equal to $2 \varepsilon$
corresponds roughly to the upper limit inferred for the UCD size.
At this force resolution only completely disrupted systems might be suitable 
UCD progenitors.
In the right panel of Figure 1 we demonstrate that the satellite 
with $c_{\rm sat}=9$ defines the upper limit of the concentration parameter 
for disruption at the timescales of interest. Note that the thick 
horizontal solid line denotes the region below which the halo may resemble 
that of a UCD galaxy. Satellites with lower values of 
concentration disrupt earlier and therefore could be associated with
UCD progenitors. These values are significantly lower than those measured
in cosmological simulations for halos in these mass scales 
(Bullock \etal 2001).

\begin{figure}
\epsscale{0.57}
\plotone{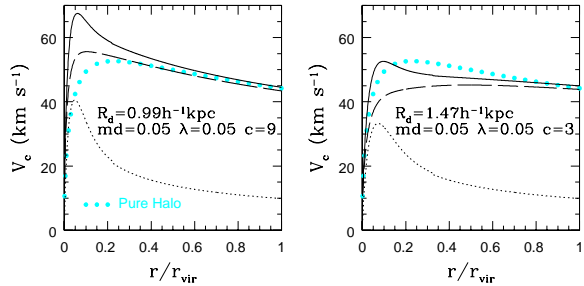}
\caption{Rotation curves of disk models with the same typical values for 
$m_{\rm d}$ and $\lambda$.
Left: A typical disk galaxy with $c_{\rm halo}=9$ has a significantly higher 
$c_{\rm eff}$ than a pure DM halo with the same concentration.
Right: The correction due to the presence of baryons yields $ c_{\rm halo}= 3$
for the starting halo concentration of this disk model.}
\end{figure}

\section{The Effect of Baryons on Satellite Survival}

The theoretical predictions of CDM models for the concentration values
are only for precollapse halos. A real galaxy would always have 
an ``effective concentration'', $c_{\rm eff}$, higher than that of a 
pure DM halo system. The baryons 
steepen the inner density profile by both adding mass to the center 
and causing the halo to adiabatically contract responding to their infall, 
increasing considerably the resilience of satellites to tidal stripping.
In Figure 2 we present the rotation curves of two disk 
models constructed using the semi-analytical modeling of 
Mo, Mao, \& White (1998) together with the rotation curve of a pure
DM halo with concentration equal to the upper limit of disruption ($c=9$)
for the adopted standard orbit. Both models have typical values of
the disk mass fraction $m_{\rm d}=0.05$ and halo spin parameter 
$\lambda=0.05$.
In the left panel is illustrated that a typical disk greatly increases the 
effective concentration of a pure DM halo making the disruption
of a satellite galaxy significantly more problematic. 
In order to achieve an effective concentration of $c_{\rm eff} \le 9$ we need 
to start from $c_{\rm halo} < 3$ for the same typical disk parameters
(right panel). This value is more than 
4 $\sigma$ lower than the theoretical predictions for the mass range of our satellites.

We are unable to explain the origin of the UCDs within the CDM models.
We shall explore further the dependence of satellite disruption on their orbital 
properties, central density slopes, and host halo structure to address this issue in 
more detail (Kazantzidis, Mayer \& Moore, in preparation).


\begin{references}
\reference Bullock, J. S., \etal 2001, \mnras, 321, 559
\reference Colpi, M., Mayer, L., \& Governato, F. 1999, \apj, 525, 720
\reference Drinkwater, M. J., Jones, J. B., Gregg, M. D., \& Phillipps, S. 2000, 
	 PASA, 17, 227
\reference Drinkwater, M. J., \etal 2003, Nature, 423, 519
\reference Ghigna, S., \etal 1998, \mnras, 300, 146
\reference Kazantzidis, S., Magorrian, J., \& Moore, B. 2003, ApJ submitted
\reference Klypin, A., Kravtsov, A. V., Valenzuela, O., \& Prada, F. 1999, 
	\apj, 522, 82
\reference Mo, H. J., Mao, S., \& White, D. M. 1998, \mnras, 295, 319
\reference Moore, B., Governato, F., Quinn, T., Stadel, J., \& Lake, G. 1998, 
\apj, 499, L5
\reference Moore, B., \etal 1999, \apj, 524, L19
\reference Stadel, J. 2001, PhD thesis, U.Washington.
\reference Taffoni, G., Mayer, L., Colpi, M., \& Governato, F. 2003, 
	\mnras, 341, 434
\reference Phillipps, S., Drinkwater, M. J., Gregg, M. D., \& Jones, J. B. 2001, 
	\apj, 560, 201
\end{references}
\end{document}